\documentclass[10pt,pre,showkeys, floatfix]{revtex4}

\usepackage{amsmath}
\usepackage{amsfonts}
\usepackage{graphicx}
\usepackage{bm}
\usepackage{amssymb}
\usepackage{mathrsfs}
\usepackage{pstricks,pst-node}
\usepackage{amsxtra}
\usepackage{bbm}
\usepackage{paralist}
\usepackage{color}

\usepackage{bigints}  

\DeclareMathAlphabet{\mathbfi}{OT1}{cmr}{bx}{it}
\DeclareMathAlphabet{\mathpzc}{OT1}{pzc}{m}{it}

\newcommand{\eqa}{\begin{eqnarray}}
\newcommand{\eeqa}{\end{eqnarray}}
\newcommand{\beq}{\begin{equation}}
\newcommand{\eeq}{\end{equation}}
\newcommand{\nn}{\nonumber}

\newcommand{\benumerate}{\begin{enumerate}}
\newcommand{\eenumerate}{\end{enumerate}}

\newcommand{\bitemize}{\begin{itemize}}

\newcommand{\eitemize}{\end{itemize}}

\newcommand{\der}[2]{\frac{\partial #1}{\partial #2}}

\newcommand{\dersec}[2]{\frac{\partial^{2} #1}{\partial #2^{2}}}

\newcommand{\dermixd}[3]{\frac{\partial^{2} #1}{\partial #2 ~\partial #3}}

\usepackage{color}

\usepackage{soul}

\begin{document}

\title{Symmetries and criticality of generalised  van der Waals models } 

\author{Francesco Giglio$^{\; 1)}$\footnote{{\tt email: francesco.giglio@glasgow.ac.uk}}, Giulio Landolfi$^{\; 2)}$\footnote{{\tt email: giulio.landolfi@le.infn.it, giulio.landolfi@unisalento.it}}, 
Luigi Martina$^{\; 2)}$\footnote{{\tt email: luigi.martina@le.infn.it, luigi.martina@unisalento.it}} and Antonio Moro$^{\; 3)}$\footnote{{\tt email: antonio.moro@northumbria.ac.uk} }
}

\affiliation{$^{1)}$ School of Mathematics and Statistics, University of Glasgow, Glasgow, UK }
\affiliation{$^{2)}$ Dipartimento di Matematica e Fisica ``Ennio De Giorgi" Universit\a`a del Salento \\
and I.N.F.N. Sezione di Lecce, via Arnesano I-73100 Lecce, Italy}
\affiliation{$^{3)}$ Department of Mathematics, Physics and Electrical Engineering, 
Northumbria University, Newcastle Upon Tyne, UK }

\date{\today}

\begin{abstract}
We consider a family of thermodynamic models such that the energy density can be expressed as an asymptotic expansion in the scale formal parameter and whose terms are  suitable functions of the volume density. We examine the possibility to construct solutions for the Maxwell thermodynamic relations relying on their symmetry properties and  deduce the  critical properties  implied in terms of the the dynamics of coexistence curves  in the space of thermodynamic variables.
 \end{abstract}

 
\maketitle

\section{Introduction}
\label{intro}

The study of  equilibrium and critical phenomena in fluids  is a widely addressed research subject.  Fluids display indeed a rich phenomenology and a number of equations of state  have been derived to effectively encode  complex   microscopic processes occurring as thermodynamical conditions vary. The celebrated van der Waals equation of state  
\beq  \left(P+\frac{a}{ v^{2}} \right)\, (v-b)\,=k_B\, T \,\, 
\label{eq vdw}
\eeq  
(where $v$ denotes the volume density, $P$ the pressure, $T$ the temperature and $k_B$ the Boltzmann constant)
has played a pivotal role in this context,  and constitutes a  paradigm  on which  the vast majority of phenomenological models 
have been  based and findings of theoretically based approaches have been assessed, see e.g.  \cite{unmix, review vdw in, hansen, rowlinson, review vdw fin}.  
Experimental investigations  indicate in fact that a single equation of state cannot  account of  thermodynamic properties and phase  transitions for all  fluid systems,  
both pure and mixtures,  under any conditions, and many empirical modifications of the van der Waals equations of state (\ref{eq vdw}) 
have been devised  for applications mainly in the context of chemical-engineering experiments \cite{soave, peng, martin, kubic, forero,zhong}. 
A number of studies have also been carried out with the aim to link phenomenological models to specific assumptions on the form of the molecular interaction potential, 
see e.g. \cite{percus, hayter,wang, yan, uribe,farzi,schmid, mangold,   nezbeda}.

From the point of view of statistical mechanics phase transitions occur in the thermodynamic regime, the limit where both volume and number of particles diverge in such a way that the density stays finite. In this regime even ``well behaved" analytical potentials may lead to the occurrence of singularities in the dependence of physical observables as functions of thermodynamic variables.  
The classical macroscopic approach, based on the principle of thermodynamics, permits a direct description of macroscopic extensive, and intensive thermodynamic variable via a set of differential equations known as Maxwell's relations. Maxwell's relations are equivalent to the existence of a free energy potential and specific assumptions on the functional form of the state functions enable one to derive the equations of state as solutions of these relations under suitable initial/boundary conditions \cite{moro, de nittis}.
 In this framework,  fluid phase transitions,  for example, can be interpreted as compressive shock waves in the profile of the state functions as they evolve in the space of thermodynamic variables \cite{moro,whitham}.  Shocks therefore arising as discontinuities in the profile of the state function correspond to the coexistence lines 
 of two phases,  see e.g.  \cite{barra, moro, de nittis}. 

 The advantage of the deployment of effective macroscopic models based of the analysis of Maxwell's relations, as  outlined above, allows for an analytical description of a general class of phenomena by-passing the complications related to the development and implementation of extensive numerical simulations of underlying statistical mechanical models.  A compelling example of the effectiveness of this approach is provided 
in \cite{giglio} where a four parameter family of generalised van der Waals models  results from
a suitable class of internal energy virial expansions. In particular, it is found that the volume density $v(x,t; \eta)$ satisfies the following nonlinear partial differential equation 
\begin{equation}
\label{veq}
\der{v}{t} + \der{}{x} \left\{ \frac{1}{c_{1} v - c_{3} k_B} \left[ 
c_{2} \, v^{2} + c_{4} \, k_B^{2} -\eta \, k_B \, \left(c_{1} \der{v}{t} + c_{2} \der{v}{x} \right) \right] \right\} = 0 
\end{equation}
with $x = P/T$ and  $t = 1/T$, while  the  $c_j$'s are structural constants entering the energy expansion. 
The quantity $\eta = 1/N$, where $N$ is the number of molecules, plays the role of the small parameter in the expasion of the internal energy.
For the particular choice $c_{1} = 0$ the model equation (\ref{veq})  (a Maxwell thermodynamic relation, see Section \ref{section model})
is identified with the well known Bateman-Burgers equation  which describes the propagation of nonlinear waves in regime of small viscosity 
\cite{bateman burgers}. One thus expect that as $\eta \to 0$ a generic (physical) volume density solution  evolves into a shock  wave   at finite $t$, 
corresponding to the occurrence of the gas-liquid phase transition \cite{moro, de nittis}. The occurrence of critical points and their configurations depend on the values of
 the structural constants $c_j$  and specific numerical values can be chosen to reproduce  isothermal curves of various models, e.g. van der Waals,  Soave-Redlich-Kwong and Peng-Robinson phenomenological models~\cite{giglio}. 

In this paper we propose a systematic study of the model equation (\ref{veq}), its solutions obtained  via the application of the Lie symmetry approach (see e.g. \cite{olver,stephani}) and their thermodynamic interpretation.
 The paper is organised as follows:  In Section \ref{section model} we introduce the model  underpinned by the equation  (\ref{veq}) and  highlight
 its main features of interest in the realm of gas thermodynamics.  
In Section \ref{sezione generatori} we present the results  entailed by the Lie-symmetry analysis of the equation. In Section \ref{discussion} we discuss  the application of results to equations of state and critical points for fluid systems.  Last section is  devoted to closing remarks.

\section{Generalised van der Waals model}
\label{section model}

The nonlinear differential equation (\ref{veq}) for the volume density $v = v(x,t; \eta)$ originates within the framework of laws of thermodynamics
by considering the thermodynamic energy balance equation  in the form $d \psi = \varepsilon d t + v d x$, 
where $\psi = t \mu$, where $\mu$ is the chemical potential and $\varepsilon = E/N$ is the internal energy per particle, being $N$ the total number of particles in the system.  In these variables, the Maxwell relation, locally equivalent to the existence of the thermodynamic potential $\psi$,   reads as
\beq
\der{v}{t} = \der{\varepsilon}{x}  \,.
\label{maxwell v t x}
\eeq
Following a suggestion in \cite{moro}, one can consider Maxwell thermodynamical equations 
within a nonlocal perturbative scheme where the expansion parameter is the inverse of the number of molecules $N$. 
 More precisely, it is assumed that the  internal energy density $\varepsilon$ admits the asymptotic expansion of the form
\begin{equation}
\varepsilon = \varepsilon_{0}(v) + \eta \, \varepsilon_{1}(v) \der{v}{x} + \eta \, \varepsilon_{2}(v) \der{v}{t} + O(\eta^{2}) + h(t)
\label{internal energy expansion}
\end{equation}
where $\eta =  1/N$ and the function $h(t)$ is an arbitrary function of its argument. 
The derivatives of the volume density in the internal energy density expansion   are introduced based on the observation that 
a number of perturbative approaches applied to the 12-6 Lennard-Jones potential \cite{hansen},  as well as more general type of potentials, 
lead to equations of state which depend on derivatives of the volume densities such as compressibility and  thermal expansion coefficient.

In \cite{giglio}, a class of internal energy virial expansions (\ref{internal energy expansion})  has been obtained by requiring that  the Maxwell thermodynamical relation
 (\ref{maxwell v t x})  is linearisable via the Cole-Hopf transformation 
\begin{equation}
\label{ch}
v(x,t; \eta) = - \eta k_B\; \der{\log \varphi(x,t; \eta)}{x} \,.
\end{equation} 
 This constraint restricts  the energy terms in Eq. (\ref{internal energy expansion}) to the following form
\beq
\varepsilon_{0} = - \frac{c_{2} v^{2} +c_4 k_B^2 }{c_{1} v - c_{3} k_B}\,\, , 
 \qquad \varepsilon_{1} = \frac{c_{2} k_B}{c_{1} v - c_{3} k_B} \,\, , 
 \qquad \varepsilon_{2} = \frac{c_{1} k_B}{c_{1} v - c_{3} k_B} \,\, , 
\label{varepsilon expansion}
\eeq
thereby implying that the Maxwell equation (\ref{maxwell v t x})  takes just the form (\ref{veq}), or, equivalently, the function $\varphi(x,t; \eta)$ satisfies the following linear partial differential equation
\begin{equation}
\label{phieq}
\eta^{2} \left (c_{1} \dermixd{\varphi}{x}{t} +c_{2} \dersec{\varphi}{x} \right) + \eta c_{3} \der{\varphi}{t} + c_{4} \varphi = 0 \,. 
\end{equation} 
Solutions to the above equation with a suitable physical initial condition  identify the partition function of the fluid characterised by the internal energy (\ref{varepsilon expansion}). 
 Fundamental properties of the associated thermodynamic system have been studied in \cite{giglio}.  In particular, the case $c_{1} \neq 0$, $c_{2}=c_{3}=0$,  $c_4/c_1=a k_B^{-2}$,  with  $a$  the  mean field parameter entering the van der Waals equation of state, has been investigated in \cite{barra}.  More specifically, based on microscopic arguments, it has been proven that the partition function for the standard van der Waals model satisfies a Klein-Gordon type equation, that is precisely Equation ~(\ref{phieq}) with  $c_{2}=c_{3}=0$ and  $c_4/c_1=a k_B^{-2}$.  
In other words, the model (\ref{internal energy expansion})-(\ref{varepsilon expansion}) can devise systems whose thermodynamics is anchored to that of van der Waals gases, with novel contributions to the effective molecular interaction  controlled by the  nonvanishing parameters $c_2$ and $c_3$.  
 Hence, although the condition of linearisability of Maxwell's relation might seem restrictive, this produces a family of models parametrised by three constants, i.e. the ratios $c_2/c_1, c_3/c_1,c_4/c_1$ with $c_1\neq 0$ , see (\ref{varepsilon expansion}).
Moreover, additional functional parameters (specifically, the co-volume) arising form the general solution of the equation (\ref{phieq}) or (\ref{veq})  can fixed via the corresponding initial conditions.

Equation (\ref{veq}) represents a viscous conservation law and as such its generic solution is expected to develop classical shock waves in finite $t$ in the inviscid limit  $\eta \to 0$, corresponding to the thermodynamic limit $N\to \infty$.  That is, in the thermodynamic regime isothermal curves can be interpreted as  nonlinear wave solutions to (\ref{veq}) that break in correspondence of the gas-liquid critical point \cite{moro,de nittis}.  Beyond the critical point solutions to the inviscid limit of the equation (\ref{veq}) obtained by setting $\eta = 0$ are in fact no longer single-valued. The region where the solution is multi-valued corresponds to the critical region where multiple phases emerge.  The underlying criticality is therefore captured by  the implicit solution written in the  \emph{hodograph} form  
\beq
x+ \varepsilon_0' (v)\, t=f(v)
	\label{eq hodograph}
	\eeq
that is a solution to the Riemann-Hopf equation 
\beq
\label{eq RH}
\der{v}{t} - \varepsilon_0' (v) \,\,\der{v}{x} = 0 \,\, ,  
\eeq  
obtained from the equation (\ref{veq}) with $\eta=0$, where $f(v)$ is an arbitrary function of  volume density and primes denote differentiation with respect to volume, i.e. 
\beq
\varepsilon_0' (v)= -\frac{c_2}{c_1} +\frac{(c_2 c_3^2+c_4 c_1^2 )}{c_1 \,(c_1 v-c_3 k_B)^2}\, k_B^2
\,\, .
\label{varepsilon prime}
\eeq
Any  particular choice of the function  $f(v)$ specify  the entropic contribution of the ensemble of molecules composing the fluid by means of the relation $f(v)=s'(v)$, where $s(v)$ stands for the  entropy density.  Evaluating the hodograph solution  \eqref{eq hodograph} at $t=0$ one has $x=f(\overline{v})$,
where $\overline{v}=f^{-1}(x)$  
can be interpreted as the volume density of the fluid when  temperature and pressure are large but such that  the ratio $x=P/T$ is finite. The hodograph function  $\Psi(v):=x +\varepsilon_0' (v)\, t - f(v)$ thus permits to write down the equation of state for the system under investigation as $\Psi(v)=0$.
Thermodynamical critical points are associated with the  \textit{critical sector}  of the Riemann-Hopf equation (\ref{eq RH}), defined by solution to 
the system of simultaneous equations \cite{kodama}  
\beq
\Psi(v)=  \Psi'(v)= \Psi''(v)=0\,\,.
\label{eq critical sector}
\eeq  
In this context, the appearance of coexistence curves  is understood as the propagation of \emph{weak solutions} to  (\ref{eq RH}) in the space of thermodynamic variables \cite{moro}, the Rankine-Hugoniot shock condition \cite{whitham} 
 being tantamount to the  Maxwell's equal areas rule.

The above framework can be applied to the  model of   interacting  molecules  described by an equation of state that, in the very high temperature regime,  reduces to  that of a system of non-interacting hard spheres, i.e. $ (\overline{v}-b) x=k_B$, where $b\in\mathbb{R} ^+$ is a parameter proportional to the volume occupied by a molecule of fluid \cite{callen}. 
It then follows that for any value of $t$ the equation of state for the system is given by the hodograph implicit solution $\Psi(v)$ with the particular choice 
\begin{equation}
\label{eos:ic}
 f^{hs}(v):=\frac{k_B}{v-b}\,\,, 
\end{equation}
and that the system  \eqref{eq critical sector}  admits the solution given by 
\begin{equation}
\label{eos:cp 2}
x_c=\frac{c_1  \left[ (27 b^2 c_1^2-54 b c_1 c_3 k_B)\,c_2 +(28 c_2 c_3^2 +c_1^2c_4)k_B^2\right]}{8 k_B  (c_2 c_3^2+c_1^2 c_4) (b c_1-c_3 k_B )} \, \,
,\quad t_c=\frac{27 c_1^2  (b c_1-c_3 k_B )}{8 k_B (c_2 c_3^2 +c_1^2c_4) } \,\, , \quad  
v_c=   3 b-\frac{2 k_B c_3  }{c_1} \, \, .
\end{equation}
The known critical point for a van der Waals fluid is recognised setting $c_{2}=c_{3}=0$ and  $c_4/c_1=a k_B^{-2}$.
All structural constants for the model $c_j$  specify the critical points. Domain restrictions arise in the form  $c_3/c_1<b/k_B$ and $c_2/c_1<0$ upon demanding that $t_c>0$ 
and the partition function be well-behaved \cite{giglio}. 
 It is also assumed that $c_2c_3^2+c_1^2 c_4\neq 0$ as when $c_2 c_3^2+c_4 c_1^2 =0$ the internal energy $\varepsilon_0$ depends on volume density linearly and the model  does not support  phase transitions. In other words the hodograph equation  (\ref{eq hodograph}) give a travelling wave solution  with constant characteristic speed,  $x-\frac{c_2}{c_1}t=f(v)$. 

To conclude, the internal energy density expansion  (\ref{internal energy expansion}) 
enables one to exploit a formal mathematical analogy between the theory of nonlinear waves and isothermal  curves for fluids.    The expansion (\ref{varepsilon expansion}) yields to the Maxwell thermodinamic equation (\ref{veq}) that is valid for a global  description of a class of fluid systems,  inside and outside the critical region. Outside the critical region, the van der Waals equation of state arises at the leading order in the expansion parameter and for the choice of parameters $c_2 = c_3 =0$ and $c_4/c_1=a k_B^{-2}$.  Inside the critical region, the analysis of solution through shock-wave techniques  provides coexistence lines consistently with Maxwell's equal areas rule.

\section{ Symmetry generators  of Equation  (\ref{veq}) }
\label{sezione generatori}

Symmetry methods are widely applied  to examine physical systems. In particular,  the adoption of a Lie group-theoretical framework is a well established way to proceed 
while  dwelling upon differential equations. Our purpose here is  to  characterise the symmetry group of the partial differential equation  (\ref{veq}), i.e. the whole class of continuous transformations acting on dependent and independent coordinates that   transforms solutions of the equation to other solutions, and to infer the possible consequences in a thermodynamic perspective. We point out that analysis is accomplished by assuming that all the  parameters $c_j$  in Equation (\ref{veq})  are  non-vanishing and, in addition, $c_2 c_3^2+c_4 c_1^2 \neq0$.

As  discussed at lenght in a number of works and textbooks, among which we mention \cite{olver, stephani}, 
the task of extracting the symmetry group of a differential equation can be implemented by looking for the admissible prolonged symmetry vector fields of the type
\beq W= g_1(x,t,v) \, \frac{\partial}{\partial x} +g_2(x,t,v) \, \frac{\partial}{\partial t}+g_3(x,t,v) \, \frac{\partial}{\partial v} \,\, .
\eeq
Since (\ref{veq}) is a second order equation, to determine the functions $g_j(x,t,v)$  that are possibly  allowed,  the infinitesimal criterion of invariance of the differential equation (\ref{veq}) $\sf{pr}^{(2)} W[\Delta]=0$ must be analysed, being $\Delta$ the l.h.s. of (\ref{veq}) and $\sf{pr}^{(2)} W$
the so-called {\em second prolongation}  of the vector field $W$ (see \cite{olver}) for details).
Since  the approach is standard and can be made efficient through symbolic mathematical  computations,  
it is sufficient for the purpose of this paper to report  directly the result.

 Assuming  that none of the coefficients involved in Equation (\ref{veq}) do vanish and $c_2 c_3^2+c_4 c_1^2 \neq0$ \cite{condizione esclusa}, 
it turns out that the sought symmetry group is generated through the action of the following three  vector fields  
\beq
W_1=
\frac{\partial}{\partial x} \,\, , \qquad 
W_2=
\frac{\partial}{\partial t} + \frac{c_2}{c_1} \frac{\partial}{\partial x}
\,\,, \qquad W_3= t \frac{\partial}{\partial t} +  \left( 2\frac{c_2}{c_1} t -  x \right) \frac{\partial}{\partial x} +
  \, \left(v-k_B\frac{c_3 }{c_1} \right)\frac{\partial}{\partial v} \,\, .
\label{W1 W2 W3}
\eeq
Each of these operators defines a one-dimensional subgroup of point-transformation, i.e. a group of local  transformations $G(\{x,t,v\}; \lambda)$ whose action on the triplet $\{x,t,v\}$ returns a new triplet of variables that depend on $\{x,t,v\}$ and on a single real parameter.
The meaning of operators $W_1$ and $W_2$ is self-evident because they are are associated with rigid translations in the $x$ and $t-\frac{c_1}{c_2} x$ directions. The corresponding one-dimensional subgroups are settled via $G_1(\{x,t,v\}; \lambda_1)=\{x+\lambda_1, t,v\}$ and   $G_2(\{x,t,v\}; \lambda_2)=\{x+\frac{c_2}{c_1} \lambda_2, t+\lambda_2,v\}$. The explicit one-parameter group of symmetry transformation implied by the generator $W_3$   is explicitly given as follows: 
\beq
G_3(\{x,t,v\};\lambda_3)=\left\{ e^{-\frac{c_1}{c_2}\lambda_3} x +2\, t\, \frac{c_2}{c_1} \, \sinh\left(\frac{c_1}{c_2}\lambda_3\right)  , \, \,
e^{\frac{c_1}{c_2}\lambda_3} t, \,\,e^{\frac{c_1}{c_2}\lambda_3} v+ k_B\,\frac{c_3}{c_1} (1-e^{\frac{c_1}{c_2}\lambda_3}) \right\}\,\,.
\eeq
A 3-parameter group of symmetry thus underlies the differential problem (\ref{veq}), the identity element being recovered by performing the limit  where the real parameters 
$\lambda_1$,$\lambda_2$ and $\lambda_3$ vanish.  It is worth to remark at this stage that the inherent question concerned with the symmetry transformations outcoming 
whenever a linear combination of symmetry generators (\ref{W1 W2 W3}) is considered finds an immediate answer.   Indeed, operators of the form $\tilde{W}\equiv W_3+\alpha_1 W_1+ \alpha_2 W_2$ with $\alpha_{1,2}$ arbitrary real constants can be actually given a structure of the type $W_3$ upon suitable shifts of the real independent variables $x$ and $t$. 
Of course, the transport of solutions into other solutions through symmetry operations can be ruled by constructing sequences of transformations as well, each of which referring to a given symmetry generator. In such a case, one expects that the order in which  different transformations are performed  in the sequence matters because while $W_1$ and $W_2$ are commuting vector fields they both do not commute with $W_3$. However, results coming by permutation of the order of transformations in a given sequence are linked each to the other by mere shifts of thermodynamical variables $x$ and $t$.  As a consequence, the lesson one ultimately learns from the analysis of Equation (\ref{veq}) by means of a group-theoretical approach is that if a function $v=v_0(x,t; \eta)$ solves Equation (\ref{veq}) then also 
\begin{equation}
v_\lambda(x,t;\eta)=e^{\frac{c_1}{c_2}\lambda_3} \,\, 
v_0(\tilde{x},\tilde{t};\eta)+ (1-e^{\frac{c_1}{c_2}\lambda_3}) \,\frac{c_3 k_B}{c_1} \,\, , 
\label{trasporto W3}
\end{equation}
with
\beq
\tilde{x} = \Lambda_1+e^{\frac{c_1}{c_2}\lambda_3} x - 2\, t\, \frac{c_2}{c_1} \sinh\left(\frac{c_1}{c_2}\lambda_3 \right) 
 \,\, ,  \qquad \tilde{t}= e^{-\frac{c_1}{c_2}\lambda_3} t +\Lambda_2 \,\,  
\label{x tilde t tilde}
\eeq
 does, being $(\lambda_3, \Lambda_1,\Lambda_2)$ a triplet of real {\em deformation} parameters. No other local symmetry transformations acting in the space $(x,t,v)$ can be devised when it is assumed that $\eta  c_1 c_2 c_3 c_4\neq 0$ along with $c_2 c_3^2+c_4 c_1^2 \neq0$.

\section{Equations of state and dynamics of critical points under symmetry transformations}
\label{discussion}
 
 Given a state of the fluid, specified by  the variables $(x,t,v)$, the above solutions \eqref{trasporto W3} and \eqref{x tilde t tilde} 
parametrised by $\Lambda_1$, $\Lambda_2$ and $\lambda_3$ describe an orbit of the associate point Lie symmetry group generating a family of equations of state.  
The action of a symmetry transformation  deforms the equations of the state  as well as the critical properties of the state functions. The approach outlined  above in Section \ref{sezione generatori}  allows to construct an infinite family of solutions to the Maxwell relation given by the equation (\ref{veq}). Hence, a separate analysis is required to select those solutions that satisfy physical assumptions to possibly capture properties of real fluids.

 In the following, we show  that in order to ensure that a solution of the group orbit satisfies the required physical properties it is necessary that the  \textit{seed solutions} $v_0$ possesses those properties,
 as for instance suitable asymptotic conditions in the thermodynamical limit.  We remark that  regardless  of the specific form of the  function $v_0(x,t; \eta)$ a symmetry transformation of the form (\ref{trasporto W3}) depending on the group parameters $\Lambda_1$, $\Lambda_2$ and $\lambda_3$ will induce a dependence of critical points on the same parameters.

 Let us denote by  $(x_c^0,t_c^0,v_c^0)$  and $(x_c^\lambda,t_c^\lambda,v_c^\lambda)$  
the triplets identifying  a critical point for the state functions $v_0$ and $v_\lambda$ respectively. As  the symmetry transformation   does not depend on $\eta$  and it is not affected therefore by the thermodynamic limit $\eta\to0$,  
  the new critical volume is determined by the right hand side of (\ref{trasporto W3}),  
\[
v_c^\lambda=e^{\frac{c_1}{c_2}\lambda_3} v_c^0 +(1-e^{\frac{c_1}{c_2}\lambda_3})\frac{c_3 k_B}{c_1},
\]
and the transformed critical pressure and temperature are given by
  \begin{align}
  \notag
x_c^\lambda&  =e^{-\frac{c_1}{c_2}\lambda_3} (x_c^0-\Lambda_{1} ) +2 \frac{c_2}{c_1}  \sinh \left(\frac{c_1}{c_2}\lambda_3\right) \, (t_c^0- \Lambda_2) \,\, ,     \\
t_c^\lambda & = e^{ \frac{c_1}{c_2}\lambda_3}   ( t_c^0   -\Lambda_2 ) \,\, ,
\label{deformed x_c  t_c}
\end{align}
 (i.e., Eq. (\ref{x tilde t tilde}) with the replacements $\tilde{x} \to x_c^0, 
\tilde{t} \to t_c^0, x \to x_c^\lambda, t \to t_c^\lambda$). 
 It is worth noting that the symmetry approach allows to study specify a family of models and their critical properties via the initial condition on the equation (\ref{veq}).

It is also   interesting to study the effect of the group transformation at the level of the partition function and Gibbs free energy potential. This step is in fact propaedeutic to  the determination of the phase-diagrams for the deformed equations of state $v_\lambda$.
As by definition we have 
\[
v_\lambda(x,t; \eta)= - \eta k_B \frac{\partial}{\partial x} \log \varphi_\lambda (x,t; \eta)
\] 
where   $\varphi_\lambda (x,t; \eta)$ denotes the solution to Equation (\ref{phieq}),  the partition function $\varphi_\lambda (x,t; \eta)$ is given, up to a constant factor, 
by \footnote{If $\varphi_0(x,t; \eta)$ solves Equation (\ref{phieq}) then 
$$
\varphi_\lambda (x,t; \eta) = \,\,  \varphi_0  \left( \tilde{x}\,, \tilde{t}; \eta \right) \,\, \exp \left\{ \frac{c_3}{\eta c_1} (1-e^{-\frac{c_1}{c_2}\lambda_3} ) \, \tilde{x}  +
2\frac{ c_2 c_3 }{c_1^2 \eta } \left[\cosh\left(\frac{c_1}{c_2}\lambda_3\right)  -1\right] \,\tilde{t} \right\}   \,\, , 
$$
where $\varphi_0  \left( \tilde{x}\,, \tilde{t} ; \eta \right)$ stands for the partition function of the original model with the  arguments $x,t$ replaced by the functions 
$\tilde{x}=\tilde{x}(x,t)$ and $\tilde{t}=\tilde{t}(x,t)$  given in  (\ref{x tilde t tilde}). By taking  $\varphi_0 (x,t; \eta)= \int _b^\infty dv \exp\left(- \frac{\Omega_0}{\eta k_B}\right) $ 
with $ \Omega_0=xv+\varepsilon_0(v)t -s_0(v) $ as in \cite{giglio}, one can  write $\varphi_\lambda (x,t; \eta) $ in the form (\ref{varphi lambda})-(\ref{def Omega}).}
 \beq
 \varphi_\lambda (x,t; \eta)= 
 \bigintsss _{b_\lambda} ^\infty dv \exp\left( -\frac{\Omega_\lambda}{\eta k_B}\right) 
 \label{varphi lambda}
 \eeq
where $\Omega_\lambda$ plays the role of the Gibbs free energy density of the form
    \beq
  \Omega_\lambda=
x v  + \varepsilon_0(v)\,t -s_\lambda(v)   \,\, .
\label{def Omega}
\eeq  
The deformed  entropy density obtained from $s_0(v)$ and associated with function $v_0(x,t)$ is given by
\beq
s_\lambda(v)=s_0\left( e^{-\frac{c_1}{c_2}\lambda_3} \left[v -   \frac{c_3}{c_1} k_B (1-e^{\frac{c_1}{c_2}\lambda_3} ) \right] \right) 
+ e^{-\frac{c_1}{c_2}\lambda_3}\left( \frac{c_2}{c_1} \Lambda_2- \Lambda_1\right) \,  v
+  \frac{(c_2c_3\,v+c_1c_4 k_B)}{c_1(c_1 v-c_3 k_B)} \, \, e^{\frac{c_1}{c_2}\lambda_3} \, \Lambda_2 k_B \,\, .
\label{s lambda}
\eeq
The quantity $b_\lambda=e^{\frac{c_1}{c_2}\lambda_3} b +\frac{k_B\,c_3}{c_1} (1-e^{\frac{c_1}{c_2}\lambda_3})$  corresponds to the minimum value of the deformed volume density. 
The domain restrictions 
$\frac{c_2}{c_1}<0$ and $\frac{c_3}{c_1}< \frac{b}{k_B}$ on the structural constants $c_j$'s ensure that   $b_\lambda$  is positive, 
 being lower than the minimum value for the seed solution $v_0$ for positive values of $\lambda_3$.

The formula (\ref{varphi lambda}) represents a smooth map providing a family of deformed partition functions starting from a seed partition function. 
The asymptotic evaluation of deformed partition functions (\ref{varphi lambda}) is then obtained standardly by Laplace's formula, implying the equation
\beq
\label{eq:eoslambda}
\Omega'_\lambda=
 x+\varepsilon'_0(v)\,t-s_\lambda'(v)= 0 \, \, , 
\eeq
where $\Omega_{\lambda}$ is given by \eqref{def Omega} with the entropic term \eqref{s lambda} and 
$\Omega'_\lambda=\frac{\partial \Omega_{\lambda}}{\partial v}$. Equation~(\ref{eq:eoslambda})
gives the possible equations of state  along  the symmetry group orbits, 
with the function $f_{\lambda}(v)\equiv s_\lambda'(v)$ assigning the entropic contribution to the equation of state. 
 Remarkably, the derivative of the potential $\Omega_\lambda$
with respect to the volume  coincides with the hodograph function 
introduced in Section \ref{section model}.

We note that the expressions (\ref{deformed x_c  t_c}) require that the liquid-to-gas critical point occurs in the physical domain $x,t\geq 0$ when  the deformation parameters satisfy the following constraints
\begin{equation}
 \label{eq:domainlambda}
\Lambda_2\leq t_c^0\,,\quad \Lambda_1\leq x^0_c -\frac{c_2}{c_1}\left( 1- e^{ \frac{2 c_1}{c_2}\lambda_3}\right) (t^0_c -\Lambda_2)\,.
\end{equation} 
A further physical constraint corresponds to the request that in the limit of zero pressure the infinite volume density is implied. 
If the starting entropy $s_0$  shapes an equation of state that meets the above physical specifications,  the function $f(v)=s'_0(v)$ goes to zero in the limit $v\to\infty$, the pressure accordingly vanishing as $t\to0$.
Under these circumstances,  by fixing $c_1 \Lambda_1 =c_2\Lambda_2$ we retrieve the desired asymptotics either for the new equation of state.

 Let us consider, for example, the case of the  van der Waals equation of state as a seed solution in the limit $\eta \to 0$ for which  the entropy $s_{0}$ is given by hard sphere formula $s^{hs}(v)=k_B \log(v-b)$. 
Hence, the corresponding deformation of the function $f^{hs}(v)=k_B (v-b)^{-1}$  (Eq. \eqref{eos:ic}) is given by
\begin{align}
f^{hs}_{\lambda}(v)
=\frac{k_B}{ v-b_\lambda} +
	 e^{-\frac{c_1}{c_2}\lambda_3}\left( \frac{c_2}{c_1} \Lambda_2- \Lambda_1\right) -\frac{k_B^2 (c_1^2c_4+c_2c_3^2)}{c_1(c_1 v-c_3k_B)^2} \,
	 e^{\frac{c_1}{c_2}\lambda_3}\,\Lambda_2 \,\, .
		\label{eq f lambda}
			\end{align}
Evaluating the deformed equation of state $x+ \varepsilon_0' (v) \, t=f^{hs}_\lambda(v)$
  as $t \to 0$ and $v \to \infty$ we have 
  \[
  x_\infty=\lim_{v \to\infty} f^{hs}_\lambda(v) = e^{-\frac{c_1 \lambda_3}{c_2}} \left(\frac{c_2}{c_1} \Lambda_2-\Lambda_1\right).
  \] 
 The condition that the pressure must vanish when volume diverges implies the constraint $c_2 \Lambda_2=c_1 \Lambda_1$. 
In the limit of infinite pressure the volume density attains its minimum value $b_\lambda$, for any choice of acceptable parameters.

\begin{figure}[h]
	\begin{center}
		\includegraphics[scale=0.42]{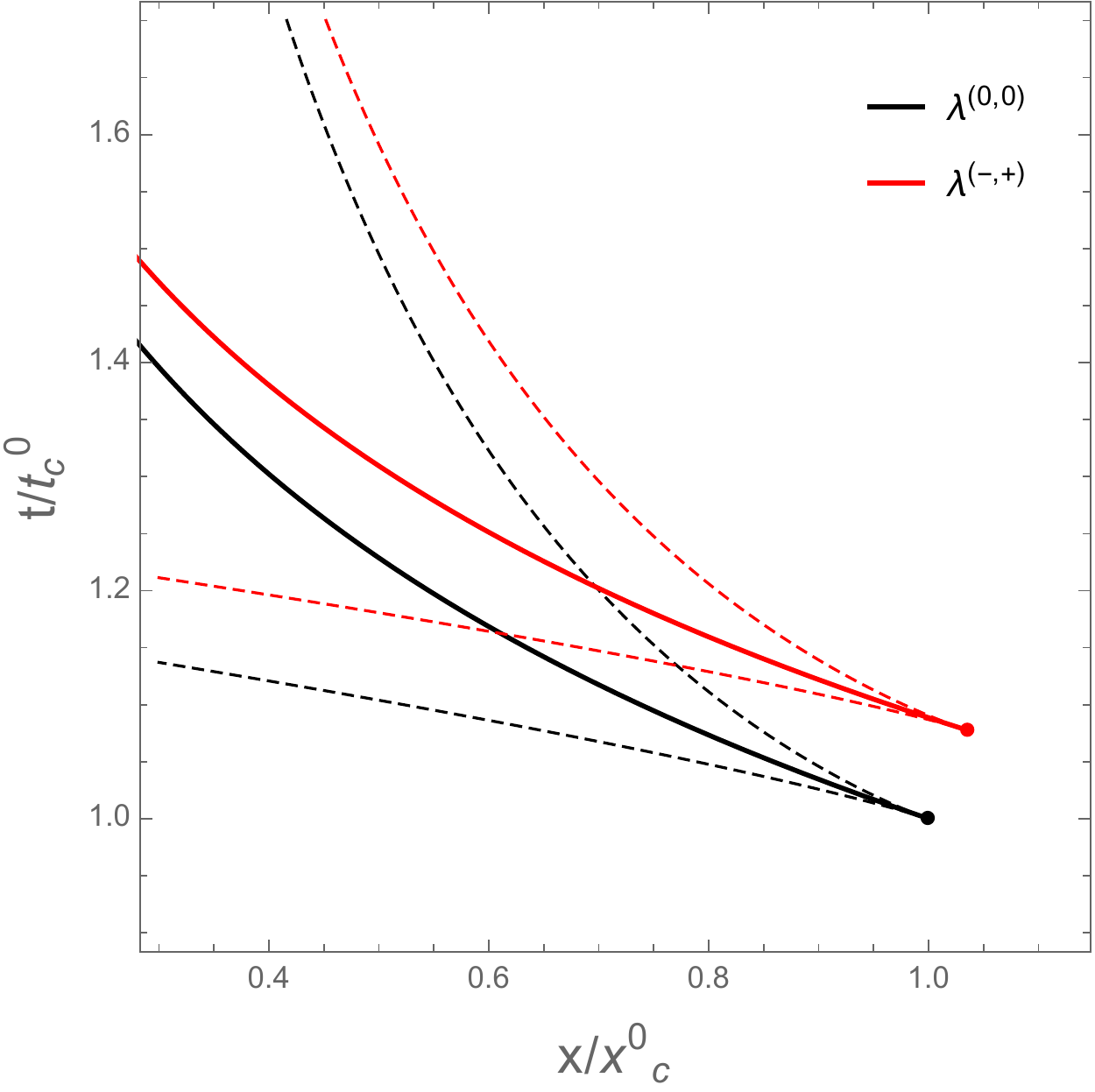}   (a) \quad
			\includegraphics[scale=0.42]{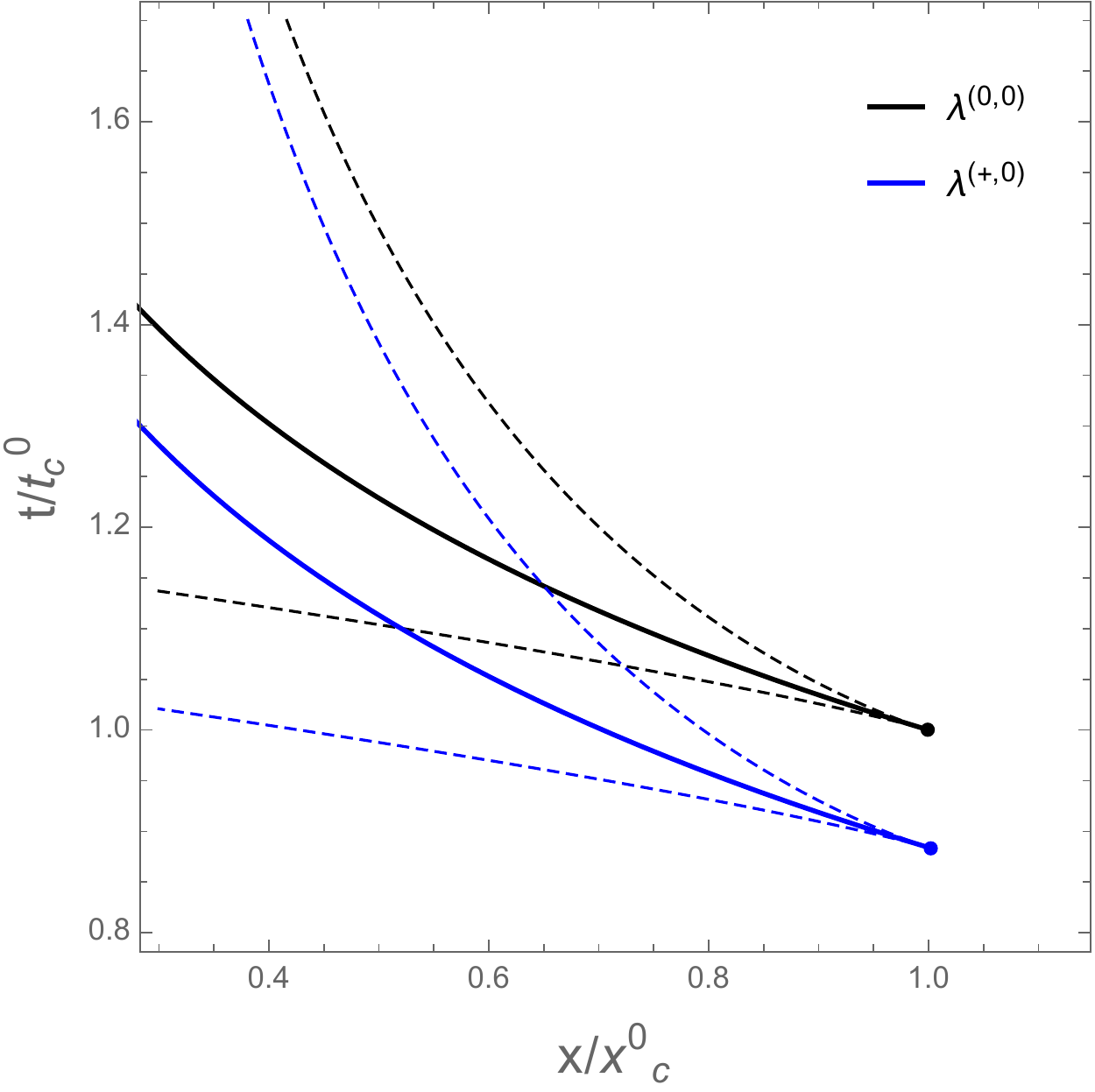}   (b)  \quad 
		\caption{
			\label{Fig:phasediagrams}
			 Examples of dynamics of shock trajectory under symmetry group action. 
			The coexistence line (solid line) originates from the critical point (solid circle) and crosses the critical region (delimited by the dashed lines).  
			Black: seed solution to (\ref{veq}) $v_0$ that captures the critical point of the Soave-Redlich-Kwong equation of state for hydrogen 
			gas  ($b= 4.419 \times10^{-29} \, m^3$,	$c_2/c_1=-8452\, \text{Pa}$, $c_3/c_1=-1.356\times 10^{-6} \, \text{K\,Pa}^{-1}$ and 
			$c_4/c_1=a k_B^{-2}=3.582\times 10^{-4} \, \text{K}^2\,\text{Pa}^{-1}$). 
			Red: new solution $v_\lambda$ gained for $\Lambda_1= \frac{c_2}{c_1} \Lambda_2$  
			and $\lambda^{(-,+)}= (\Lambda_2=-0.005 \, \text{K}^{-1},\lambda_3=300 \, \text{Pa})$.
			Blue:  $v_\lambda$ when $\Lambda_1= \frac{c_2}{c_1} \Lambda_2$ and $\lambda^{(-,0)}=(\Lambda_2=-0.005  \, \text{K}^{-1} , \lambda_3=0)$.
			For the sake of homogeneous  schematics, thereinafter legend $\lambda^{(0,0)}$ is used for the $v_0$'s to stress that the case is concerned 
			with no deformation, $\Lambda_1=\Lambda_2=\lambda_3=0$. 
}
	\end{center}
\end{figure}
We now study the phase diagrams associated to the deformed equations of state.
Figure \ref{Fig:phasediagrams} shows the coexistence line corresponding to the shock trajectory and the critical region enclosed by the {\em general fold}  \cite{arnold} of the equation, 
i.e. the solutions of the system  $\Psi_{\lambda}(v)=\Psi^{'}_{\lambda}(v)=0$ for the hodograph function  $\Psi_{\lambda} (v):=x+ \varepsilon_0'(v) t - f^{hs}_{\lambda}(v)$. 
The chosen numerical values  for the structural parameters  are such that the seed solution $v_0$ to equation (\ref{veq}) provides the critical point of the Sove-Redlich-Kwong equation of state.  The figure illustrates the effect of the deformation of this solution induced by the symmetry 
transformation (\ref{trasporto W3})-(\ref{x tilde t tilde})   as the parameters $ \Lambda_2$ and $\lambda_3$ vary such that the  constraint 
$c_{1 }\Lambda_1 = c_2 \Lambda_2 $ is fulfilled 
in terms of the resulting displacement of the critical points and the coexistence curve in the plane of thermodynamical variables $(x,t)$.
 The coexistence curve gives the  trajectory of the shock emerging from the critical point as specified
by the asymptotic evaluation of the integral (\ref{varphi lambda}) for $t>t_c$. More specifically, for small $\eta$ we have 
\[
\varphi_{\lambda}(x,t;\eta) \simeq \sum_k \sqrt{\frac{2 \pi \,\eta \,k_B}{\Omega''_{\lambda,k}} } \,e^{-\frac{\Omega_{\lambda,k}}{\eta \, k_B}},
\] 
where the sum index $k$ runs over the local minima at fixed pressure and temperature of the Gibbs free energy  (\ref{def Omega}). 
Along the coexistence lines we have  for all pairs of solutions associated to gas and the liquid phases 
 \[
 \Omega_{\lambda,j}=\Omega_{\lambda,k}.
 \]


\begin{figure}[h!]
	\begin{center}
	
	\includegraphics[scale=0.45]{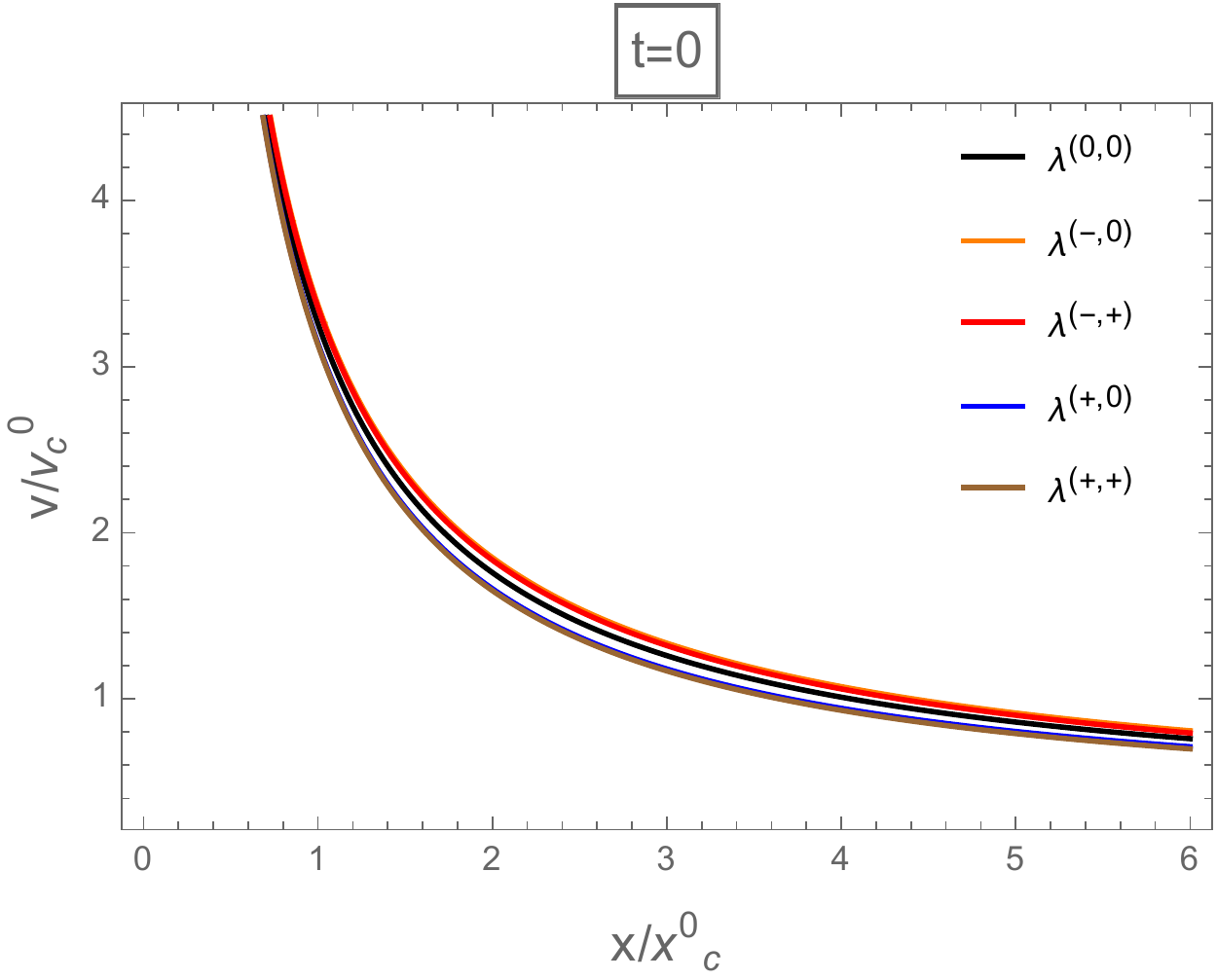}   (a) \quad
	\includegraphics[scale=0.45]{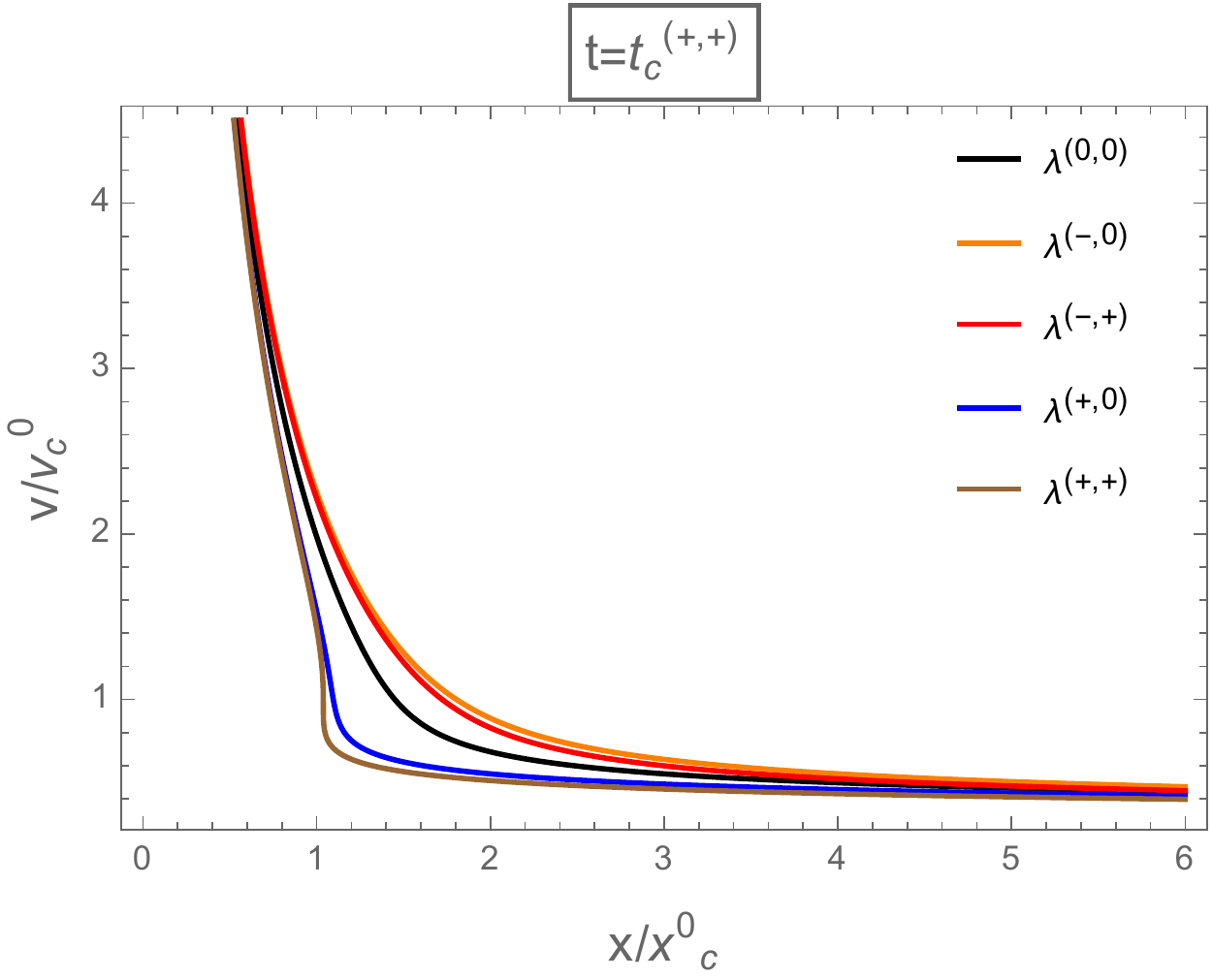}   (b)  \quad \\
	\includegraphics[scale=0.45]{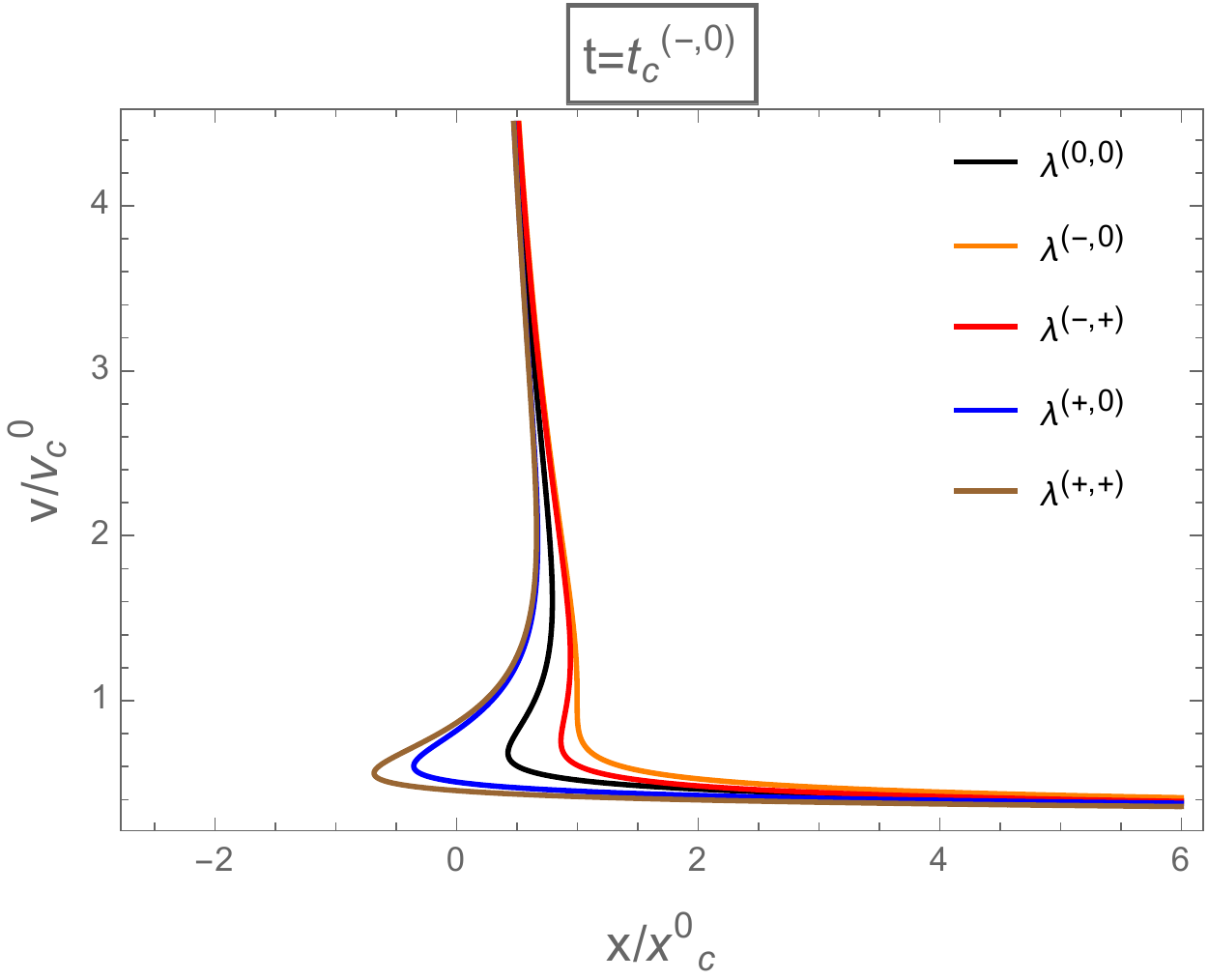}   (c) \quad
		\includegraphics[scale=0.45]{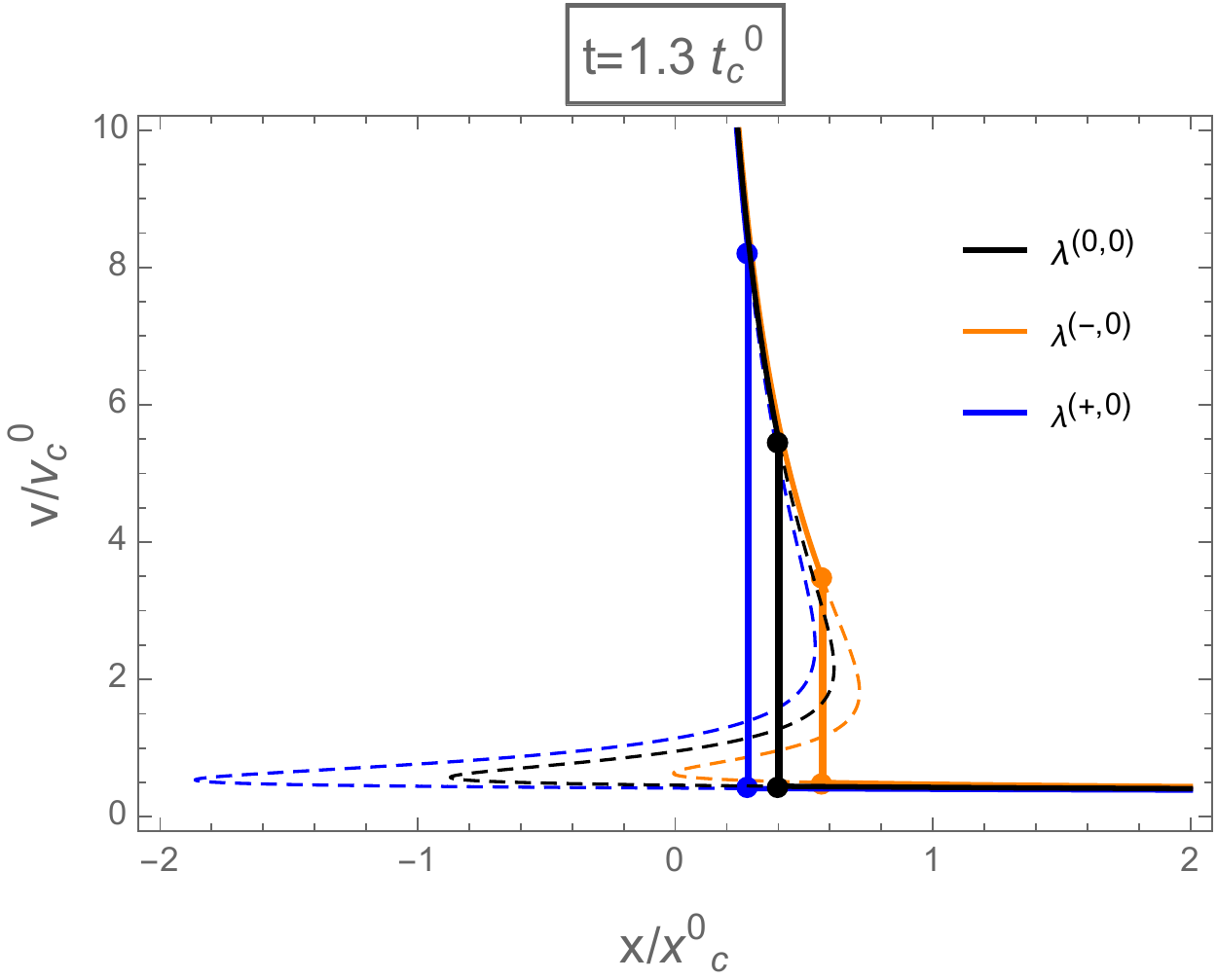}   (d)
		\caption{
			\label{Fig:isothermal}
Dynamics of isothermal curves for the solution $v_0$ (black) that captures the critical point of the Soave-Redlich-Kwong equation of state for hydrogen gas and new equations of 	state  $v_\lambda$ (brown/blue/red/orange) obtained through nontrivial pairs $(\Lambda_2,\lambda_3)$ jointly with the ansatz $\Lambda_1=  \Lambda_2 c_2 / c_1$.  
The plots refer to the pairs of deformation parameters  $\lambda^{(+,+)}=(0.005 \text{ K}^{-1},\, 300 \text{ Pa})$ (brown),  $\lambda^{(+,0)}=(0.005 \text{ K}^{-1},\, 0 \text{ Pa})$ (blue),   $\lambda^{(-,+)}=(-0.005 \text{ K}^{-1},\, 300 \text{ Pa})$ (red) and $\lambda^{(-,0)}=(-0.005 \text{ K}^{-1},\, 0 \text{ Pa})$ (orange).
(a)  Isothermal curves at $t=0$, corresponding to infinite temperature. (b)   Isothermal curves  at $t_c^{(+,+)}\approx 0.85274 \, t_c^0$, i.e. at the critical temperature of solution $v_\lambda$ obtained with parameters $\lambda^{(+,+)}$. (c)  Isothermal curves 
at the critical temperature of equation of state $v_\lambda$ with the choice $\lambda^{(-,0)}$,  $t_c^{(-,0)} \approx 1.11645 \, t_c^{0}$. 
(d) Physical  isothermal curves all beyond their corresponding critical temperature:		
solid vertical lines are located in correspondence of the shock positions and allow to retrieve the single-valuedness of the solution by replacing the oscillatory behaviour (dashed curves) with a finite jump in the volume density.   
}
	\end{center}
\end{figure}
The effect of deformations  on the isothermal curves is illustrated in Figure \ref{Fig:isothermal}  where we chose four different non-trivial sets of deformation parameters $(\Lambda_1,\Lambda_2,\lambda_3)$.
Figure \ref{Fig:isothermal}(a) shows that in the limit of infinite temperature  $t\to0$, isotherms associated to the equation of state  for $v_\lambda$'s depart from the ones associated to $v_0$  but overall preserve their the qualitative features.  
As the temperature decreases isothermal curves develop a multivalued behaviour as functions of $x$. However, the curves differ quantitatively as each solutions evolves at a different characteristic speed, which explains the shift in the position of the shock profile. In particular, for fixed symmetry parameters, there is a temperature interval where the two solutions $v_0$ and $v_\lambda$ exhibit opposed behaviours. Indeed, as illustrated in Figure \ref{Fig:isothermal}(b),  the volume density  $v_{\lambda}$ associated with the pair $(\Lambda_2,\lambda_3)$  from set  $\lambda^{(+,+)}$  develops a gradient catastrophe at $t=t_c^{(+,+)}$,   
whilst the seed solution and the solutions corresponding to the choice of parameters $\lambda^{(+,0)}$, $\lambda^{(-,+)}$ and $\lambda^{(-,0)}$ 
remain single-valued.
 For temperatures below the critical value,  the multivalued volume density profile $v_\lambda$ with $\lambda^{(+,+)}$ is replaced by classical shock. 
 At $t=t_c^0$, it is the seed solution $v_0$ that experiences a gradient catastrophe, while, for example, the solutions $v_\lambda$ constructed with negative $\Lambda_2$ are single valued. 
At sufficiently low temperatures single-valuedness is lost for all the five solutions (see Fig.  \ref{Fig:isothermal}(c)), 
and physical isotherms are given by  shock waves traveling towards lower pressures. An illustrative example of the application of the shock fitting procedure to determine coexistence lines at fixed temperature is explicitly demonstrated in Figure  \ref{Fig:isothermal}(d). 
 In this respect, it is important  to 
note how solutions $v_\lambda$ differ at low temperature. In this regime the contribution  from the internal energy is dominant over the entropic one.
However, evidently different realisations of the phase transition result from minor differences for the entropic term $s_\lambda'(v)$, which is 
the initial datum for the nonlinear differential equation governing the equation of state   and  establishes the fluid's behaviour far from the critical region.

 \section{Conclusions }
 
We identified all the Lie-point symmetries generated by infinitesimal operators giving finite group transformations 
for a thermodynamic model, introduced recently \cite{giglio}, based on the differential equation (\ref{veq}) for the volume density.
 The model provides the first derivation of a new extension of the van der Waals model as studied in \cite{moro} valid in the critical region
and  it has proved to be effective in the construction of a new interpolating model compatible with empirical models, such as the Peng-Robinson and the 
Soave-Redlich-Kwong. We found that a rather restricted set 
of point symmetries generators underlies the  differential problem (\ref{veq})
 with all four nonvanishing structural constants $c_j$, and such that $c_2 c_3^2+c_4 c_1^2 \neq0$  (a  necessary condition for the generation of critical points).  
The class of point symmetries obtained express invariance under translations and scalings, and a linear mixing of the independent thermodynamical variables $x=P/T$ and $t=1/T$. 
The equation \eqref{trasporto W3} gives a solution depending on three real parameters  providing continuous deformations of  solutions to the differential equation  (\ref{veq}). 
Tuning the symmetry parameters permits to interpolate between already existing models for real fluid matching  qualitative properties and critical points. Critical points obtained from deformed equations of state depend on the parameters realising the action of the symmetry group and are connected to the critical points of the seed solution $v_0(x,t)$ to which   (\ref{trasporto W3}) is applied, Eqs. (\ref{deformed x_c  t_c}).
We have observed that deformations induced by the action of the symmetry group allow to model significant deviations from the seed solution in the vicinity of the critical point consistently with the behaviour of the van der Waals model at high temperature.

The problem of constructing suitable partial differential equations for state functions of thermodynamic systems and the study of critical properties in terms of critical asympotics of the solutions to these equations is an active field of research which brought further insights on a variety of classical systems, see e.g. \cite{lorenzoni 2019, dematteis,choquard,barra 2, ABLMP,BMPS}, and appears to be promising  for the study of complex systems \cite{BM,BDM}.  Studies similar to the present work can be put forward therefore for other systems of physical interest. 
Natural developments  include the study of composite systems,  such as fluid mixtures \cite{sengers}, and nematic fluids \cite{degennes}.

\acknowledgments
F.G. and A.M. acknowledge the hospitality of the  Lecce's division of I.N.F.N. and of the {\em Department of Mathematics and Physics "Ennio De Giorgi"} of the University of Salento.
 G.L. acknowledges the hospitality of the {\em Department of Mathematics, Physics and Electrical Engineering} of Northumbria University.  A.M. is supported by the Leverhulme Trust Research Project Grant RPG 2017-228. A.M. is also grateful to the London Mathematical Society, the Royal Society International Exchanges Grant IES-R2-170116 (PI A.M.), GNFM - Gruppo Nazionale per la Fisica Matematica, INdAM (Istituto Nazionale di Alta Matematica) for supporting activities that contributed to the research reported in this paper.


\begin{thebibliography}{99}
\bibitem{condizione esclusa} When $c_2 c_3^2+c_4 c_1^2 =0$ 
infinite symmetries come into play: in addition to three generators  of the type  (\ref{W1 W2 W3}), the family of symmetry generators  
 \beq 
W^\infty= \left[ G_1+\frac{c_1}{c_2} F_1(t)\right] \,  \frac{\partial}{\partial t}+\left[   F_2\left(-\frac{c_1}{c_2} x+t \right)
   +F_1(t) \right] \, \frac{\partial}{\partial x} +
   \frac{c_1}{c_2} \left( v\;+\;\sigma \frac{c_3}{c_1} \right) F_2' \left(-\frac{c_1}{c_2} x+t\right)\, 
   \frac{\partial}{\partial _{v} } \nn  
   \label{W infty}
   \eeq
is found, being    $F_1, \;F_2$  arbitrary functions of their argument and  $G_1$ constant.
\end{thebibliography}
 \end{document}